 \documentclass[lettersize,journal]{IEEEtran}
\IEEEoverridecommandlockouts
\def\BibTeX{{\rm B\kern-.05em{\sc i\kern-.025em b}\kern-.08em
    T\kern-.1667em\lower.7ex\hbox{E}\kern-.125emX}}

\usepackage{xcolor, color}
\IEEEoverridecommandlockouts

\def\BibTeX{{\rm B\kern-.05em{\sc i\kern-.025em b}\kern-.08em
    T\kern-.1667em\lower.7ex\hbox{E}\kern-.125emX}}

\usepackage{mathtools}
\usepackage[nodisplayskipstretch]{setspace}
\usepackage[utf8x]{inputenc}

\DeclareMathOperator*{\argmax}{arg\,max}

\newcommand\blue[1]{{\color{blue} #1}}

\newcommand\red[1]{{\color{red} #1}}
\usepackage{color}

\usepackage{subfigure}
\usepackage{moreverb}
\usepackage{epsfig}
\usepackage{epstopdf}
\usepackage{amsmath,amssymb,bm,amsthm,mathrsfs,dsfont}
\usepackage{fancyhdr}
\usepackage{adjustbox,lipsum}
\usepackage[font={small}]{caption}
\usepackage{color,soul}
\usepackage{graphicx}
\usepackage{amsfonts}
\usepackage{cite}
\usepackage{microtype}
\usepackage{graphicx}
\usepackage[colorlinks,bookmarksopen,bookmarksnumbered,citecolor=blue,urlcolor=red]{hyperref}

\usepackage{nomencl}
\makenomenclature

\usepackage{algorithmicx}
\usepackage{algpseudocode}
\usepackage[norelsize, linesnumbered, ruled, vlined, lined, boxed, commentsnumbered]{algorithm2e}

\newtheorem{remark}{Remark}

\allowdisplaybreaks
\usepackage{url}
\usepackage[nodisplayskipstretch]{setspace}

\begin{document}
 \setlength{\abovedisplayskip}{3.4pt}
 \setlength{\belowdisplayskip}{3.4pt}
  \setlength{\abovecaptionskip}{0.1cm}

\title{ Data-Free Knowledge Distillation for LiDAR-Aided Beam Tracking in MmWave Systems}
\author{Abolfazl~Zakeri,~\textit{Member, IEEE}, Nhan~Thanh~Nguyen,~\textit{Member, IEEE},
 Ahmed Alkhateeb,~\textit{Senior Member, IEEE},
 and Markku~Juntti,~\textit{Fellow, IEEE}
 \vspace{-3 em }
 \thanks{
 Copyright (c) 20xx IEEE. Personal use of this material is permitted. However, permission to use this material for any other purposes must be obtained from the IEEE by sending a request to pubs-permissions@ieee.org.
 \\
 A. Zakeri, N. T. Nguyen, and M. Juntti are with the Centre for Wireless
Communications (CWC), University of Oulu, Oulu 90014, Finland, Emails:
\{abolfazl.zakeri; nhan.nguyen; markku.juntti\}@oulu.fi. A.~Alkhateeb is with the School of Electrical, Computer, and Energy Engineering, Arizona State University, Email: alkhateeb@asu.edu.\\
This work was supported by the Research Council of Finland through 6G Flagship Program (grant 369116) and projects DIRECTION (grant 354901), DYNAMICS (grant 24305016), and CHIST-ERA PASSIONATE (grant 359817), by Business Finland, Keysight, MediaTek, Siemens, Ekahau, and Verkotan via project 6GLearn, and in part by the HORIZON-JU-SNS-2023 project INSTINCT (101139161). 
 }
}
\maketitle
\begin{abstract}
We propose a data-free knowledge distillation (DF-KD) framework for LiDAR-aided mmWave beam tracking, where the objective 
is to predict the optimal current and future beams from a 
sequence of past LiDAR measurements.
Specifically, we propose a knowledge inversion approach where a generator synthesizes LiDAR-like sequences 
from random noise, using a metadata loss to align the teacher's internal feature statistics of synthetic and real data, without access to raw LiDAR samples. 
The student model is then trained exclusively on the synthetic data using either the 
Kullback-Leibler (KL) divergence loss or a proposed mean 
squared error (MSE) loss between the teacher's and student's raw output logits. Simulation results on the DeepSense dataset demonstrate the effectiveness of the proposed approach. 
In particular, the proposed convolutional neural network-gated recurrent 
unit (CNN-GRU) teacher architecture yields superior DF-KD student performance compared to GRU-only alternatives, and the MSE loss achieves performance comparable to the standard 
KD loss while requiring fewer hyperparameters.
\end{abstract}
\section{Introduction} 
Beam training has long been a critical task for the performance of massive multiple-input multiple-output (mMIMO) communications. This becomes more crucial in millimeter-wave (mmWave) systems, where communication links are highly sensitive to path loss and blockages. Leveraging multimodal sensory data, such as visual, light detection and ranging~(LiDAR), and radar measurements, for communication tasks, referred to as \textit{multimodal sensing-aided communications}, has emerged as a promising approach to enhance beam training~\cite{Ahmed_deepsense,multimodality_mmwave_mag,R_Heath_mag,latin_amr_RH}. It can substantially reduce beam training overhead~\cite{RHeath_BT_multiuser,Ahmed_seman_dis} and improve beam alignment in connected vehicle environments~\cite{multimodality_connected_vechile}. The benefits are particularly pronounced in high-mobility scenarios, where proactive line-of-sight link prediction and future beam selection become critical for maintaining reliable~connectivity.
\\\indent
Building on this premise, multimodal sensing-aided beamforming has attracted significant attention recently ~\cite{zakeri_globe25sensing,Debbah_25_llm,Ahmed_LiDar_COML, Batool_modalsensing,
DL_vtc24,salehihikouei2024leveraging,Ahmed_vision,Khaled_25,RHeath_BT_multiuser,Walid_ML25}.
For instance,  Patel~\textit{et al.} \cite{RHeath_BT_multiuser} showed sensor-aided multimodal deep learning can efficiently estimate beamspace channels for multi-user mmWave MIMO, enabling interference-free beamforming and significantly improving spectral efficiency.
Jiang \textit{et al.}~\cite{Ahmed_LiDar_COML} demonstrated that LiDAR-aided machine learning (ML) can accurately predict and track optimal mmWave beams in real-world vehicular scenarios, noticeably reducing beam training overhead; this approach was further extended to multimodal-aided beam prediction in vehicular networks in~\cite{Ahmed_vision,DL_vtc24} and recently to integrated sensing and communications~\cite{Khaled_25}.
\\\indent
Most existing works rely on ML to map multimodal sensory data to optimal beam indices. However, these approaches are often limited by high training costs, model complexity, and substantial dataset requirements. 
While Park et al.~\cite{Walid_ML25} proposed a knowledge 
distillation (KD) framework to transfer knowledge from a complex multimodal 
model to a lightweight monomodal network, their approach 
still depends on large training datasets that may be 
unavailable in practice due to storage constraints at the 
UE or privacy limitations at the BS.
To address this, we propose a data-free KD (DF-KD)~\cite{DF_org} framework for LiDAR-aided beam 
tracking that eliminates the need for training data after the initial teacher training phase, while producing a low-complexity model.
\\\indent 
We build upon the beam tracking problem of~\cite{Ahmed_LiDar_COML}, where the objective is to map 
a sequence of past LiDAR measurements to the optimal current and future beam indices. The considered scenario 
involves vehicle-to-infrastructure communication with a single BS equipped 
with a uniform linear array and a LiDAR sensor, and a single-antenna mobile UE.
The developed DF-KD framework is based on knowledge inversion. Specifically, a generator synthesizes 
LiDAR-like input sequences from a random noise vector, 
guided by a metadata loss that minimizes the discrepancy between the teacher’s internal feature representations elicited by synthetic and real data, respectively. 
The student model is then trained exclusively on this synthetic data using either 
the Kullback-Leibler (KL) divergence loss or a proposed 
mean squared error (MSE) loss between the teacher's and student's raw output logits.
Simulation results on the DeepSense dataset demonstrate the effectiveness of the proposed approach. 
Specifically, we show that (i) the metadata loss is 
essential for generating informative synthetic data that 
faithfully reflects the teacher's internal feature 
distribution, (ii) the proposed convolutional neural network-gated recurrent 
unit (CNN-GRU) teacher architecture yields superior DF-KD student performance compared to GRU-only alternatives,
and (iii) the MSE loss achieves performance comparable to the 
standard KD loss while requiring fewer hyperparameters.
\\\indent 
Compared to~the closest work~\cite{Ahmed_LiDar_COML}, our work differs in three key aspects. 
First, we introduce a CNN-GRU teacher architecture that incorporates convolutional layers for spatial feature 
extraction from each LiDAR frame, yielding a more powerful and expressive teacher model. 
Second, the proposed DF-KD framework eliminates the need for real LiDAR data after teacher training by using synthetic data generated via knowledge inversion.
Third, the resulting student model achieves an approximately $23$ times reduction in the number of parameters compared to \cite{Ahmed_LiDar_COML}, significantly reducing model complexity and enabling deployment in resource-constrained~platforms.
\section{System Model and Problem Formulation}
 \subsection{System Model}
 We consider a mmWave communications system with a BS, located at a fixed position, and a mobile UE.
The BS is equipped with a uniform linear array (ULA) with $N$ antennas, and the UE is equipped with a single-antenna receiver. 
Moreover, the BS is also equipped with a LiDAR sensor to sense the environment and use the sensing information for communications. 
\\\indent
At time slot $t=1,2,\dots$, the BS transmits the data signals to the UE using the beamforming vector ${\mathbf{w}(t)\in\mathbb{C}^{N\times 1} }$. 
We consider a codebook-based beamforming design, i.e., ${\mathbf{w}(t)\in \mathcal{W}}$, where $\mathcal{W}=\{\mathbf{w}_1,\dots, \mathbf{w}_M\}$ is the codebook of all possible beamforming vectors with size $M$. Using analog beamsteering, the beamforming vectors in $\mathcal{W}$ have constant modulus, i.e., each vector satisfies $\|\mathbf{w}_m\|_{2}^{2} = 1,~m\in\{1,\dots,M\}$.
\\\indent
Denote $\mathbf{h}(t)\in\mathbb{C}^{N\times 1} $ as the channel vector between the BS and the UE at slot $t$. We adopt a block-fading channel model and express the channel at slot $t$ as 
\begin{align}
    \mathbf{h}(t) = \sum_{c=1}^{ C } \alpha_c(t) \mathbf{a}(\theta_c(t),\phi_c(t)),
\end{align}
where $C$, $\alpha_c(t)$, $\theta_c(t)$, and $\phi_c(t)$ are, respectively, the number of paths,  complex gain, azimuth angle, and elevation angle of the $c$-th path at time $t$. Furthermore, let ${s}(t)\in\mathbb{C}$ be the transmit (data) signal to the UE with $\Bbb{E}\{|{{s}(t)}|^2\}=1$.
The received signal by the UE is then given~by 
\begin{align}
    y(t) 
    &= \mathbf{h}^{\textsf{H}}(t) \mathbf{w}(t){s}(t) + n(t),
\end{align}
where  $n(t)\in \Bbb{C}$ is additive white Gaussian noise following the distribution $\mathcal{CN}(0,\sigma^2)$, with $\sigma^2$ denoting the noise variance at the UE's receiver.
\\\indent
Given a predefined codebook, the beamforming vector $\mathbf{w}(t)\in\mathcal{W}$ is specified by the beam index $m(t)\in\{1,\dots,M\}$ at time $t$.
The beamforming problem is then formulated as finding the optimal beam index $m^{\star}(t)$ that maximizes the received SNR of the UE, i.e., 
\begin{align}\label{op_main1}
    m^\star(t) = \argmax_{m(t)\in\{1,\dots,M\}}~  |\mathbf{h}^{\mathsf{H}}(t)\mathbf{w}_{m(t)}|^2. 
\end{align}
We remark that if the channel vector $\mathbf{h}(t)$ is available, the above problem can be easily solved. In fact, the optimal beamforming is the matched filter, i.e., $\mathbf{w}(t) =\sqrt{P}\dfrac{\mathbf{h}(t)}{\|\mathbf{h}(t)\|}$, with $P$ denoting the transmit power budget, which is commonly known as a maximum ratio transmission beamformer. However, the main challenge here is that obtaining the channel information at all times requires excessive signaling overhead, additional latency, and resource consumption. 
To overcome this, similarly to,~e.g.,~\cite{R_Heath_mag,latin_amr_RH,Ahmed_vision}, we propose the idea of using LiDAR sensory data with ML for beamforming design. 
The detailed learning task is elaborated next.
\begin{remark}
    We consider LiDAR because (1) it provides superior spatial 3D resolution and (2) its emergence and adaptation in autonomous vehicles.
     Nevertheless, LiDAR incurs substantial data collection and storage overhead, making large-scale training datasets difficult and costly to maintain, particularly at the UE, or at the BS, thereby further motivating our DF-KD approach for LiDAR-aided beam tracking.
\end{remark}
It is worthwhile to note that the proposed framework is evaluated on DeepSense Scenario~8,
which predominantly contains line-of-sight (LoS) mmWave links. At mmWave
frequencies, non-LoS (NLoS) conditions introduce severe signal attenuation
that significantly degrades link quality, representing a fundamental limitation of physical propagation.
Moreover, physical obstructions may obscure the LiDAR sensing range, thereby degrading the quality of the sensory input.
Proactive blockage prediction~\cite{lidar_block_Ahmed} and LoS/NLoS-aware curriculum training~\cite{lidar_nlos_gunduz} have been proposed to mitigate
these limitations. Extending the proposed DF-KD framework
to environments with prevalent NLoS conditions constitutes an important
direction for future~work.
\subsection{ML Beamforming Problem Definition}
The goal is to use LiDAR data to perform \textit{beam tracking}, i.e., predict the current and future best beams.  
To formulate this, let $\mathbf{x}(t)\in\mathbb{R}^D$ denote the (preprocessed) LiDAR data, with fixed-length feature of size $D$, captured at the BS at time~$t$. The sequence of $L\in\mathbb{N}$ most recent LiDAR measurement data is denoted as $\mathcal{X}(t;L)=\{\mathbf{x}(t-L+1),\dots,\mathbf{x}(t)\}$. 
Considering the time-dependent nature of UE mobility 
we utilize past LiDAR data rather than relying solely on current observations from slot~$t$. Therefore, the goal is to learn a mapping function that takes the LiDAR input sequence, $\mathcal{X}(t; L)$, and predicts the optimal current/future beam indices~using~ML.
\\\indent 
\textbf{Beam Tracking Problem:} Let  $f_\mathbf{\Theta}(\cdot)$ be a mapping function, parameterized by an ML model (i.e., the neural network), $\mathbf{\Theta}$, that maps the input LiDAR data to a beam index for the current time $t$ and 
$V\in\mathbb{N}$ number of future beams $t+1,\dots,t+V$
according to
\begin{align}
    \hat{m}(t+v) = \argmax_{m \in \{1, \dots, M\}} f_\mathbf{\Theta}(\mathcal{X}(t;L);v)_m,
\end{align}
where $f_\mathbf{\Theta}(\mathcal{X}(t;L);v) \in \mathbb{R}^M, v\in\{0,1,\dots, V\}$ is the output logits for each beam~$m$ of time $t+v$, and $\hat{m}(t')$ is the predicted beam index at time $t'=t,t+1,\dots,t+V$, by the model.\footnote{This is the same as using the maximum likelihood where the probability distribution over the beam class is obtained by applying softmax to the logits.} The goal is to train the model to learn optimal $f^{\star}_\mathbf{\Theta}(\cdot)$ corresponding to $\mathbf{\Theta}^\star$ such that $\hat{m}(t') = m^\star(t')$.
\\\indent
The final beam selection for each time $t$ is generally based on top-$K$ promising beams obtained as above from the ML model. Given the top-$K$ beams, the BS can either 1) directly choose the top-$1$ beam and mitigate the beam training overhead, or 2) choose the $K$ beams and perform an over-the-air beam training only for those $K$  beams~\cite{Ahmed_LiDar_COML} with significantly reduced overhead compared to conducting an exhaustive search over the entire codebook. Next, we briefly introduce DF-KD and describe how we adapt it for the beam training task.

\section{Data-Free KD For Beam Tracking}
\subsection{Main Ideas} DF-KD builds upon conventional KD but does not require 
access to the original training dataset~\cite{DF_org}. In standard KD~\cite{Hinton_KD}, the objective is 
to transfer knowledge from a large, high-capacity teacher model $\mathsf{T}$ to a compact student model $\mathsf{S}$ 
with fewer parameters or a simpler architecture, while 
retaining as much performance as possible. This is 
typically achieved by minimizing the divergence between 
the student's output distribution and the soft output 
probabilities of the teacher, obtained by applying a 
temperature parameter $T$ to soften the logits. Beyond 
output probabilities, KD can also transfer knowledge at 
different representational levels, such as intermediate 
hidden features or relational structures between samples~\cite{KD_survey25}.
\\\indent 
In practice, training data may be unavailable due to privacy constraints, storage limitations, or data 
collection overhead. In such cases, synthetic data guided by the teacher's internal representations can enable 
knowledge transfer without access to real training 
samples. Motivated by this, we propose a DF-KD framework 
that uses only the teacher model and its internal feature 
statistics--without requiring any raw LiDAR inputs or 
ground-truth labels--to train a synthetic data generator 
and subsequently a compact student model. The overall 
procedure is summarized in Algorithm~\ref{alg_dfkd} and 
proceeds in three steps. In Step~(0), the teacher 
$\mathsf{T}$ is trained on the real LiDAR dataset 
$\mathcal{D}$, and metadata statistics are collected 
from its internal representations. In Step~(1), the 
generator $\mathsf{G}$ is trained via knowledge inversion 
using the frozen teacher and the collected metadata. In 
Step~(2), the student $\mathsf{S}$ is trained exclusively 
on the synthetic data produced by $\mathsf{G}$, with 
knowledge distilled from $\mathsf{T}$. The details of each step are described in the following~subsections.
\begin{algorithm}[t!]
   \caption{\small Overall Data-Free KD Algorithm}
   \label{alg_dfkd}
   \SetKwInOut{Input}{Initialize}
   \SetKwInOut{Output}{Output}
   \SetKwComment{Comment}{/* }{ */}
   \setlength{\AlCapSkip}{1em}
   \Input{
       Dataset $\mathcal{D}$; randomly initialized, teacher model~$\mathsf{T}$, generator 
       model~$\mathsf{G}$ and student~model~$\mathsf{S}$
   }
   
\Comment{Step (0): Train teacher and collect metadata}
Train teacher model $\mathsf{T}$ on the training split of 
$\mathcal{D}$ via cross-entropy minimization

\For{each batch $x$ in the training split of $\mathcal{D}$}{
    Extract the last hidden layer output of $\mathsf{T}(x)$ 
    and append to metadata set $\mathcal{M}$
}
Compute global statistics: $\mu_\mathcal{M} \gets 
\mathrm{mean}(\mathcal{M})$,\; 
$\sigma^2_\mathcal{M} \gets \mathrm{var}(\mathcal{M})$

\Comment{Step (1): Train generator via knowledge inversion}
\For{each epoch}{
    Sample noise vector for each batch: 
    $\tilde{x} \sim \mathcal{N}(0, \mathbf{I})$
    
    Generate synthetic samples: $\hat{x} \gets \mathsf{G}(\tilde{x})$
    
    Compute teacher hidden features $f_\mathsf{T}$ from 
    $\mathsf{T}(\hat{x})$

    Compute metadata loss: $\mathcal{L}^{\mathsf{G}}_{\text{meta}} 
    \gets \|\mu_{f_\mathsf{T}} - \mu_{\mathcal{M}}\|^2_2 + 
    \|\sigma^2_{f_\mathsf{T}} - \sigma^2_{\mathcal{M}}\|^2_2$
    
    Update generator parameters (i.e., backpropagation)
}

\Comment{Step (2): Train the student using synthetic data}
\For{each epoch}{
    Sample noise vector for each batch: 
    $\tilde{x} \sim \mathcal{N}(0, \mathbf{I})$
    
    Generate synthetic samples: $\hat{x} \gets \mathsf{G}(\tilde{x})$
    
    Get teacher logits: $z_\mathsf{T} \gets \mathsf{T}(\hat{x})$
    
    Get student logits: $z_\mathsf{S} \gets \mathsf{S}(\hat{x})$
    
    Compute the loss either by~\eqref{eq_student_kl_loss} 
    or~\eqref{eq_student_mseloss_DF}
    
    Update student parameters (i.e., backpropagation)
}
\Output{Trained student model $\mathsf{S}$}
\end{algorithm}
\begin{figure}[t!]
    \centering
    \includegraphics[width=0.75\linewidth]{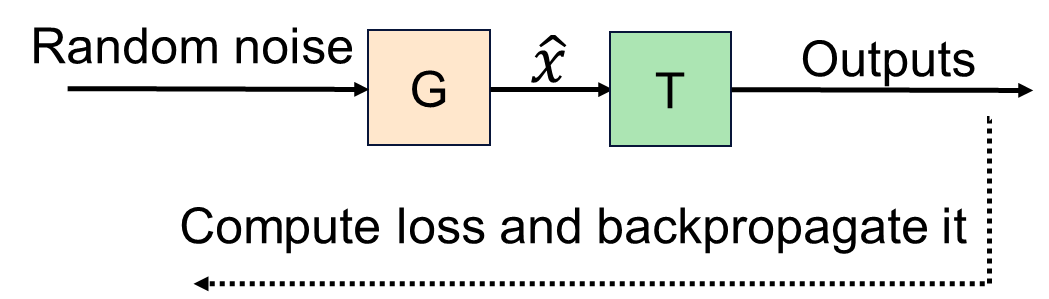}
    \caption{A schematic of training the generator $\mathsf{G}$ using the pretrained teacher $\mathsf{T}$}
    \label{fig_generator_train}
    \vspace{-2 em}
\end{figure}
\subsection{Synthetic Data Generation} To enable knowledge transfer without access to the original LiDAR 
training data, we propose a knowledge inversion framework in which 
a generator model $\mathsf{G}$ is trained to produce synthetic 
LiDAR-like sequences that elicit similar internal representations 
in the teacher as real training data would, as schematically 
depicted in Fig.~\ref{fig_generator_train}. The procedure consists 
of two phases: metadata collection and generator training.

\textbf{Metadata Collection:} After the teacher $\mathsf{T}$ is 
fully trained on $\mathcal{D}$, a single forward pass is performed 
over all $N$ training samples in evaluation mode. For each sample, 
the final hidden state of the GRU layer, 
$\mathbf{h} \in \mathbb{R}^{d}$, is extracted and concatenated 
across all training samples to form the metadata matrix 
${\mathcal{M} \in \mathbb{R}^{N \times d} }$. Importantly, 
$\mathcal{M}$ does not contain any LiDAR data nor the ground-truth beam 
labels, only the teacher's internal representations, making it 
a compact and privacy-preserving statistical summary of the 
training distribution.

\textbf{Generator Architecture:} The generator $\mathsf{G}$ is a two-layer multi-layer perceptron (MLP) that maps a random noise vector 
$\tilde{x}\sim\mathcal{N}(0, \mathbf{I})$ to a synthetic LiDAR 
sequence, with rectified linear unit (ReLU) activations in both layers. The output is 
reshaped to match the real LiDAR input format of $L$ frames each 
with $D$ features, and zero-padded with $V$ frames to produce the 
full input tensor required by the~teacher.

\textbf{Generator Loss:} The generator is trained by minimizing the discrepancy between the statistical distribution of the 
teacher's internal representations elicited by synthetic data 
and those encoded in $\mathcal{M}$. Specifically, for each batch 
of synthetic inputs $\hat{x} = \mathsf{G}(\tilde{x})$, the frozen teacher produces a batch of hidden features 
$f_\mathsf{T} \in \mathbb{R}^{B \times d}$. The metadata loss is 
defined as:
\begin{equation}
    \mathcal{L}^{\mathsf{G}}_{\text{meta}} = 
    \left\|\mu_{f_\mathsf{T}} - \mu_{\mathcal{M}}\right\|^2_2 + 
    \left\|\sigma^2_{f_\mathsf{T}} - 
    \sigma^2_{\mathcal{M}}\right\|^2_2,
    \label{eq_meta_loss}
\end{equation}
where $\mu_{f_\mathsf{T}}$ and $\sigma^2_{f_\mathsf{T}}$ are the mean and variance of the synthetic features in the current batch, 
and $\mu_{\mathcal{M}}$ and $\sigma^2_{\mathcal{M}}$ are the global mean and variance computed from $\mathcal{M}$. This formulation 
encourages the generator to synthesize LiDAR-like sequences whose distribution at the teacher's internal representation layer matches that induced by real training data, without requiring access to any raw LiDAR samples during generator training.
\subsection{Training the Student Through Distillation}
\begin{figure}[t!]
    \centering
    \includegraphics[width=0.95\linewidth]{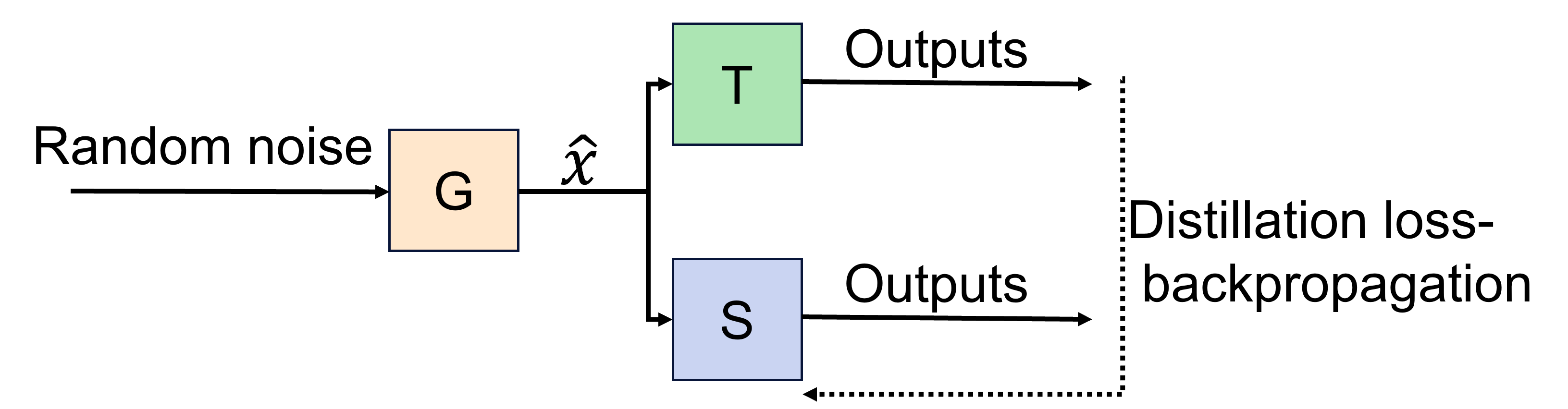}
    \caption{A schematic of training the student $\mathsf{S}$ using the pretrained teacher $\mathsf{T}$ and pretrained generator $\mathsf{G}$}
    \label{fig_DFKD}
   \vspace{-2 em}
\end{figure}

Once the generator is trained, it is used to produce synthetic 
data for performing KD using the standard procedures, as schematically depicted in Fig.~\ref{fig_DFKD}. However, since no ground-truth labels are available for the synthetic data, the 
cross-entropy loss, as in standard KD~\cite{Hinton_KD}, 
cannot be applied. Instead, the KL divergence 
loss between the teacher's and student's output distributions 
remains a valid and effective choice. Additionally, inspired by~\cite{understanding_kd}, we also consider an MSE loss 
between the teacher's and student's logits as an alternative. 
We numerically compare the performance of the KL and MSE losses in Section~\ref{sec_numres}.

\textbf{KL Divergence Loss:} KL divergence measures how one 
probability distribution diverges from another. In KD, it aligns 
the student's output probability distribution with the teacher's. 
Given the teacher and student logits softened by a temperature 
parameter $T$, the KL divergence between the resulting softmax 
distributions is computed as
\begin{equation}
    \mathcal{L}_{\text{KL}} = \frac{T^2}{B} \sum_{j=1}^{B} 
    \sum_{i=1}^{M} p_{i,j} \log \left( \frac{p_{i,j}}{q_{i,j}} 
    \right), \label{eq_student_kl_loss}
\end{equation}
where $p_{i,j} = \text{softmax}(z^{\mathsf{T}}_{i,j}/T)$ is the 
teacher's soft probability for beam $i$ in batch $j$ and 
$q_{i,j} = \text{softmax}(z^{\mathsf{S}}_{i,j}/T)$ is the 
student's soft probability for beam $i$ in batch $j$. Note that 
a higher temperature $T$ softens the distributions, allowing the 
student to better capture the relative class similarities learned 
by the teacher. The $T^2$ factor compensates for the reduction 
in gradient magnitude introduced by the temperature scaling.

\textbf{MSE Loss Between Logits:} We further consider an MSE 
loss that directly matches the raw logits (pre-softmax outputs) 
of the teacher and the student, defined as
\begin{equation}
    \mathcal{L}_{\text{MSE}} = \frac{1}{B \cdot M} \sum_{j=1}^{B} 
    \sum_{i=1}^{M} \left( z^{\mathsf{S}}_{j,i} - z^{\mathsf{T}}_{j,i} 
    \right)^2. \label{eq_student_mseloss_DF}
\end{equation}
This loss encourages the student to replicate the structure and 
scale of the teacher's output distribution without relying on 
label information. Furthermore, unlike the KL loss, it does not 
necessitate tuning of the temperature parameter $T$.

\section{Numerical Results}\label{sec_numres} 
This section presents simulation results to evaluate the performance 
of the proposed DF-KD framework and to examine the impact of different loss functions on the generator and student training.\footnote{The implementation code is available at \href{https://github.com/AZakeri94/DF-KD-for-LiDAR-aided-Beam-Tracking.git}{GitHub}.} 
The simulation scenario and parameters are described next.

\textbf{Scenario:} As in~\cite{Ahmed_LiDar_COML}, we consider vehicular-to-infrastructure (V2I) communication using Scenario~8 
from the DeepSense~6G dataset~\cite{Ahmed_deepsense}, where the vehicle speed ranges from 3 to 18~km/h.\footnote{We note that the 3--18~km/h speed range reflects the characteristics of DeepSense~6G Scenario~8, which may correspond to a typical urban vehicular setting as in, e.g.,~[10]. The only other LiDAR-assisted V2I scenario (Scenario~9) provides a similar mobility profile, while the higher-speed scenarios in the dataset correspond to vehicle-to-vehicle configurations without a fixed LiDAR-equipped BS and are therefore not directly compatible with the considered V2I framework. Evaluating the proposed DF-KD framework under broader mobility conditions remains an important direction for future work.}
The moving vehicle carries an 
omnidirectional mmWave 60~GHz transmitter to communicate with the BS. The BS is equipped with a 60~GHz receiver that has a 
16-element phased array and uses a predefined beam codebook of 
$M = 64$ beams. Further details about the dataset and the scenario 
can be found in~\cite{Ahmed_deepsense, Ahmed_LiDar_COML}.

\textbf{Dataset Construction:} The DeepSense Scenario~8 dataset consists of multiple measurement sequences, each 
comprising consecutive LiDAR frames and corresponding beam indices. The dataset is partitioned at the sequence level 
into training ($70\%$), validation ($15\%$), and test 
($15\%$) splits, preserving the chronological order of 
measurements to reflect realistic deployment conditions. 
Within each split, sliding windows of $L$ consecutive LiDAR frames are extracted with a stride of one, and each 
window is paired with the corresponding current and $V$ 
future beam indices as prediction targets, yielding a total 
input sequence length of $L + V$ time slots. The training loader uses a batch size of 64 with shuffling, 
while the validation and test loaders use a batch size of 
128 without shuffling.

\textbf{Model and Parameters:} We propose a convolutional neural network-gated recurrent unit (CNN-GRU) architecture for the teacher model, compared to a pure GRU model in~\cite{Ahmed_LiDar_COML}, where two strided \texttt{Conv1d} layers (16 and 32 filters, kernel size 5, stride 2) first extract spatial features 
from each LiDAR frame, followed by a fully connected embedding layer 
projecting to a 64-dimensional representation, and a single-layer 
GRU with hidden size 128. The student model is a lightweight GRU 
with embedding size 16 and hidden size 24. 

The observation window is $L = 8$ and the future 
prediction length is $V = 3$, so the model predicts the current and three future beam slots.
 The generator is a two-layer MLP with hidden size 128, taking a random noise 
vector of length $d_z = 500$ as input and producing a synthetic 
LiDAR sequence of $L = 8$ frames, each with $D = 216$ features, 
followed by zero-padding of length $V = 3$ to match the teacher 
input format.  For all models, the optimizer is Adam with a learning rate $10^{-3}$ and a weight decay $10^{-4}$, and a 
cosine annealing learning rate scheduler is used. 
The generator is trained for $500$ epochs with gradient clipping (norm equals $1.0$). 
Following generator training, $5{,}000$ synthetic LiDAR sequences are generated and fixed, and the DF-KD student is trained on this fixed synthetic dataset for $500$ epochs with early stopping (patience equals $10$) on the validation loss. 
The teacher, vanilla student, i.e., student without KD, and standard KD student are each trained for up to $100$ 
epochs with early stopping (patience equals $10$).
For the performance criteria, Top-1 and Top-5 accuracy are shown.
To ensure statistical reliability, the dataset construction 
and all model training procedures are repeated across $20$ independent Monte Carlo runs with fixed pre-specified seeds, and the standard error of the mean 
($\pm$SEM) is also reported across runs in all figures.

\begin{figure}[t!]
\hspace{-0.45cm}
\subfigure[Top-1] 
{
\includegraphics[width=0.24\textwidth]{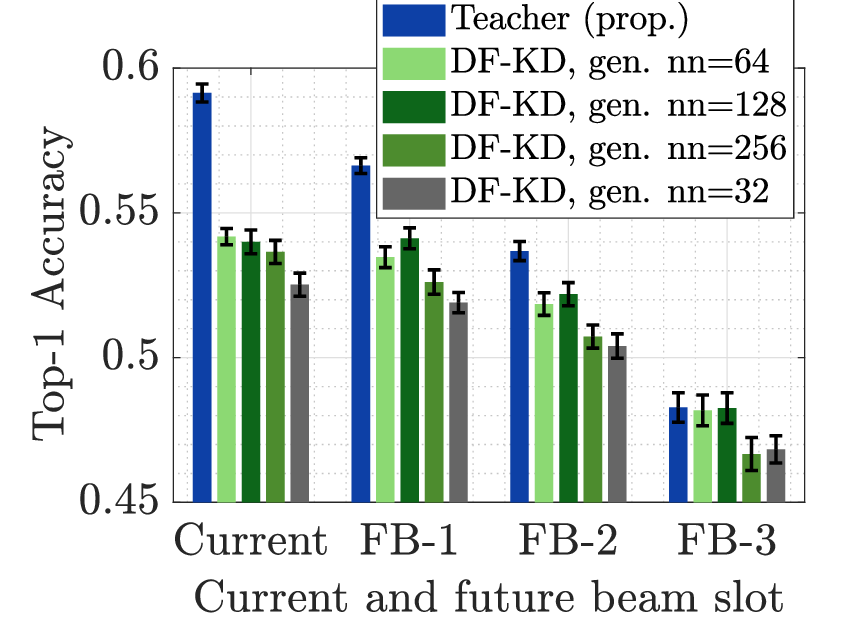}
\label{fig_gnrimpact_t1}
}
\hspace{-0.45cm}
\subfigure[Top-5]{
\includegraphics[width=0.25\textwidth]{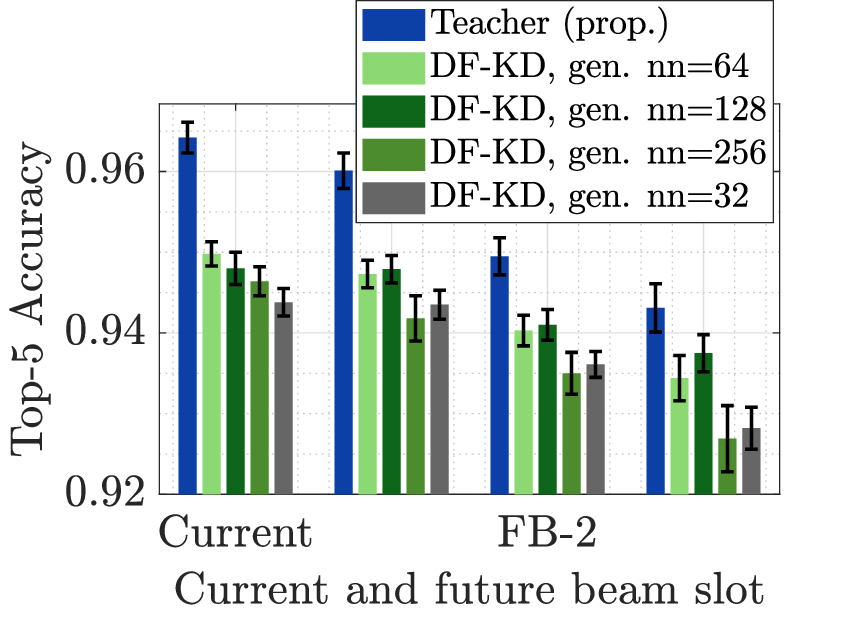}
\label{fig_gnrimpact_t5}
}
\caption{Top-1 and Top-5 (beam tracking) accuracies of the proposed teacher and the DF-KD student for different generator hidden layer~sizes} 
\label{fig_gnrimpact_}
\vspace{-4 mm}
\end{figure}

\textit{Impact of the generator capacity:} Fig.~\ref{fig_gnrimpact_} presents the {Top-1} and Top-5 
accuracies of the proposed teacher and the DF-KD student 
trained with generators of different hidden layer sizes 
(32, 64, 128, and 256 neurons), using the KL distillation 
loss with temperature $T = 2$. The results confirm that 
generator capacity has a meaningful impact on DF-KD 
performance. A generator with 32 hidden neurons is 
insufficient to capture the teacher's feature 
distribution, yielding the lowest DF-KD accuracy across 
all prediction slots. Increasing the hidden size to 128 neurons consistently achieves the best DF-KD performance 
in both Top-1 and Top-5 accuracy, striking a balance 
between expressive capacity and generalization. A generator with 64 neurons performs comparably, while increasing to 256 neurons worsens performance, 
particularly at farther future beam slots. This is likely due 
to overfitting to specific teacher activation patterns 
and resulting in low-diversity synthetic samples. Based 
on these results, a generator with 128 hidden neurons 
is adopted in all subsequent experiments.

\begin{figure}[t!]
\hspace{-0.45cm}
\subfigure[Top-1] 
{
\includegraphics[width=0.25\textwidth]{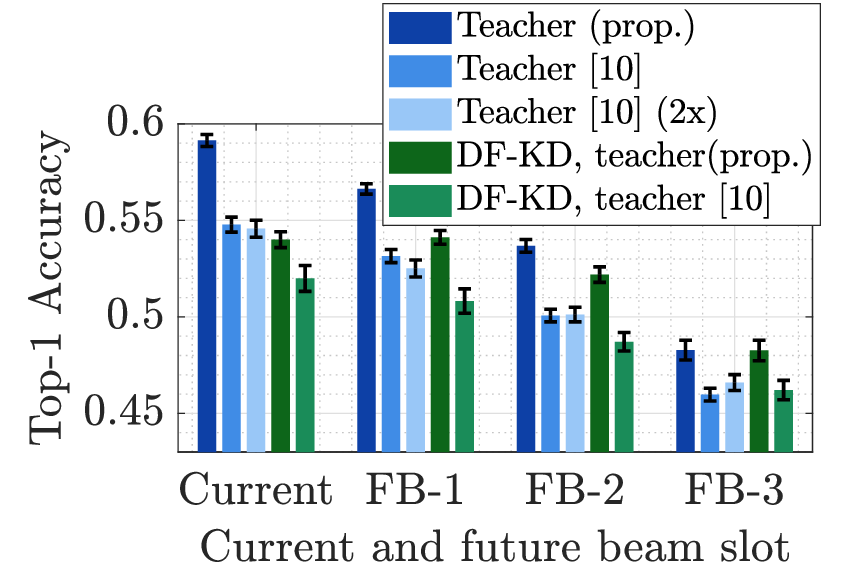}
\label{fig_Timpact_t1}
}
\hspace{-0.45cm}
\subfigure[Top-5]{
\includegraphics[width=0.24\textwidth]{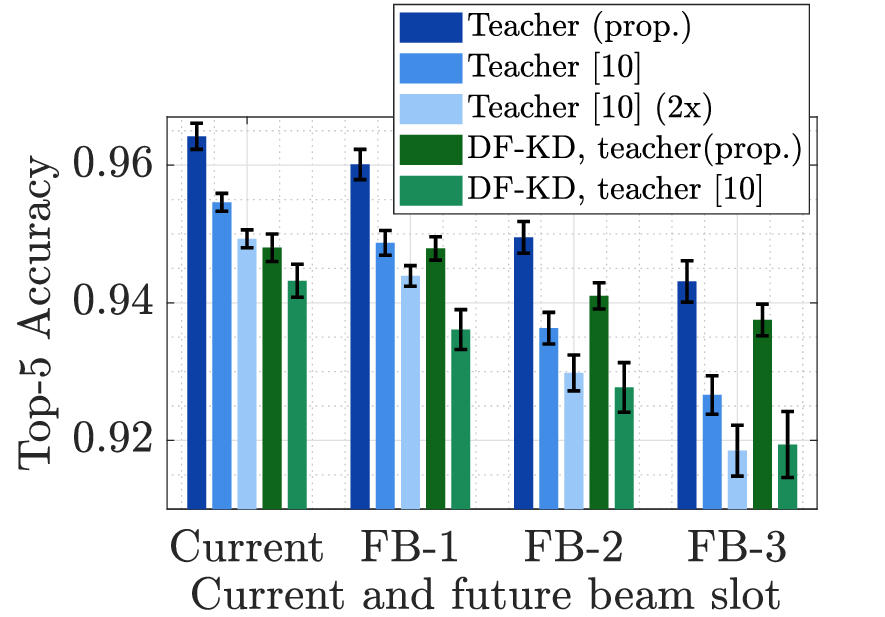}
\label{fig_Timpact_t5}
}
\caption{Top-1 and Top-5 beam tracking accuracies of the proposed 
CNN-GRU teacher and the DF-KD student compared against the 
GRU-based teacher of~\cite{Ahmed_LiDar_COML} and its scaled-up 
variant} 
\label{fig_Timpact_}
\vspace{-3 mm}
\end{figure}

\textit{Impact of the teacher architecture:} Fig.~\ref{fig_Timpact_} compares the Top-1 and Top-5 beam tracking accuracies of three 
teacher architectures: the GRU-based model from~\cite{Ahmed_LiDar_COML} 
(Teacher~[10]), a scaled-up version with twice the hidden and 
embedding dimensions, i.e., Teacher~[10]~(2$\times$), and the proposed 
CNN-GRU teacher (Teacher~(prop.)). The DF-KD student performance under each teacher is also shown. The proposed CNN-GRU teacher 
achieves the highest accuracy across all prediction slots in both 
metrics, with a Top-1 current-slot accuracy of $59.1\%$ compared 
to $54.8\%$ and $54.6\%$ for Teacher~[10] and Teacher~[10] (2$\times$), respectively. Notably, simply scaling up the GRU architecture does not yield consistent improvements, 
confirming that the CNN-based spatial feature extractor provides 
a qualitatively better inductive bias for processing LiDAR frames 
than increasing GRU capacity alone. Furthermore, the DF-KD student 
trained under the proposed teacher outperforms those trained under 
both GRU-based teachers across all slots, demonstrating that a stronger and more expressive teacher translates directly into 
better student performance under the data-free constraint.

\begin{figure}[t!]
\hspace{-0.45cm}
\subfigure[Top-1] 
{
\includegraphics[width=0.25\textwidth]{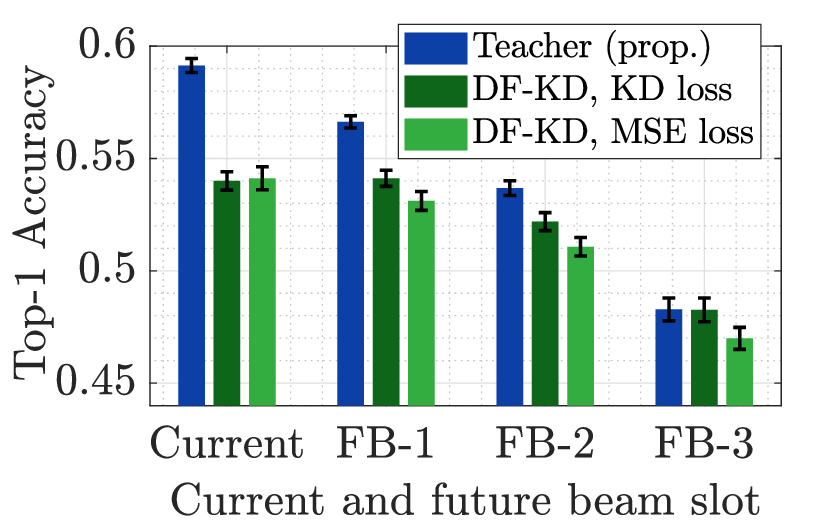}
\label{fig_Limpact_t1}
}
\hspace{-0.45cm}\subfigure[Top-5]{
\includegraphics[width=0.25\textwidth]{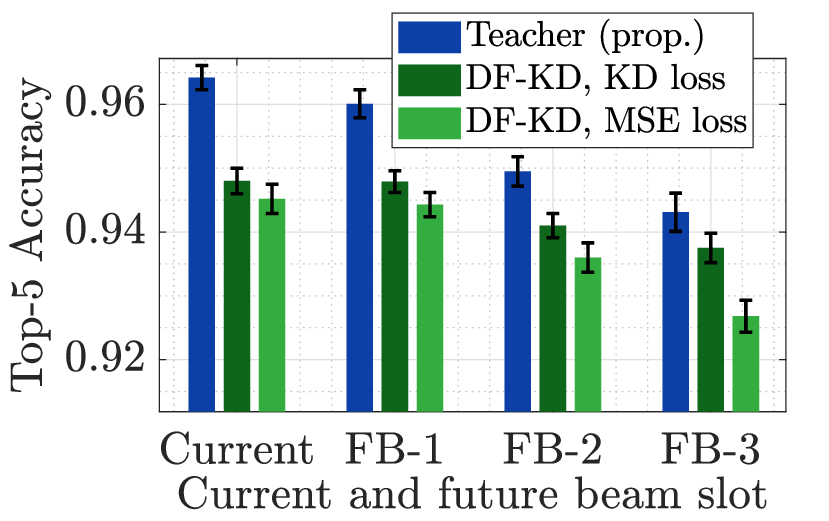 }
\label{fig_Limpact_t5}
}
\caption{ Top-1 and Top-5 accuracies of the (proposed) teacher and the DF-KD student for different student distillation losses.} 
\label{fig_Limpact_}
\vspace{-3 mm}
\end{figure}

\textit{Impact of the student distillation loss:} Fig.~\ref{fig_Limpact_} 
examines the effect of the KL divergence loss~\eqref{eq_student_kl_loss} 
and the MSE loss~\eqref{eq_student_mseloss_DF} on the Top-1 and Top-5 DF-KD student accuracy, using the proposed CNN-GRU teacher 
and the 128-neuron generator. Both losses enable the student to 
effectively distill knowledge from the teacher's synthetic outputs. 
The KL loss achieves slightly higher Top-1 accuracy at the current 
slot, while the two losses are broadly comparable across future beam slots. This suggests that the MSE loss can serve as a simpler alternative 
to the KL loss, eliminating the need for temperature hyperparameter tuning in DF-KD, and, additionally, the weighting coefficient 
$\alpha$ in standard KD~\eqref{eq_std_kd}, while preserving comparable beam tracking~performance.

\begin{figure}[t!]
\hspace{-0.45cm}
\subfigure[Top-1] 
{
\includegraphics[width=0.25\textwidth]{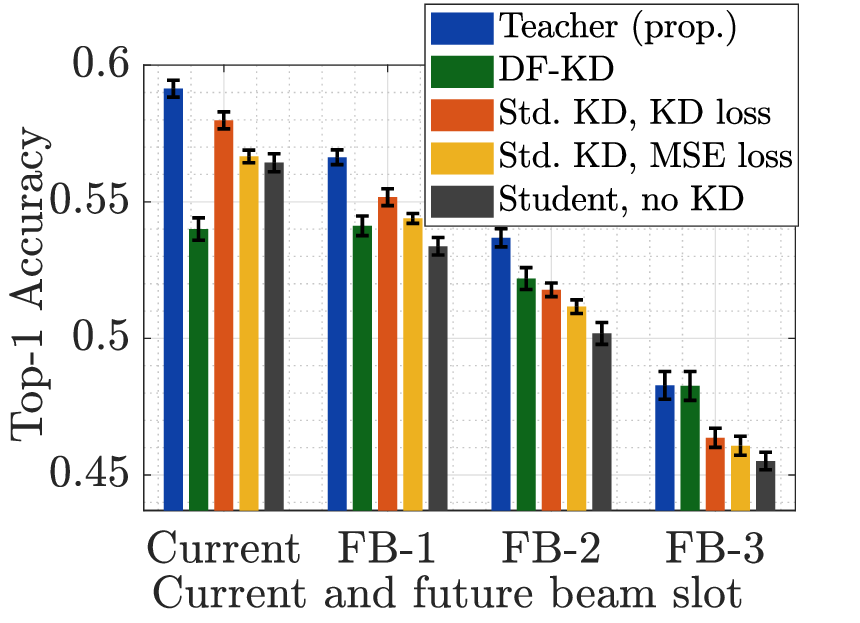}
\label{fig_dfvskd_top1}
}
\hspace{-0.45cm}\subfigure[Top-5]{
\includegraphics[width=0.25\textwidth]{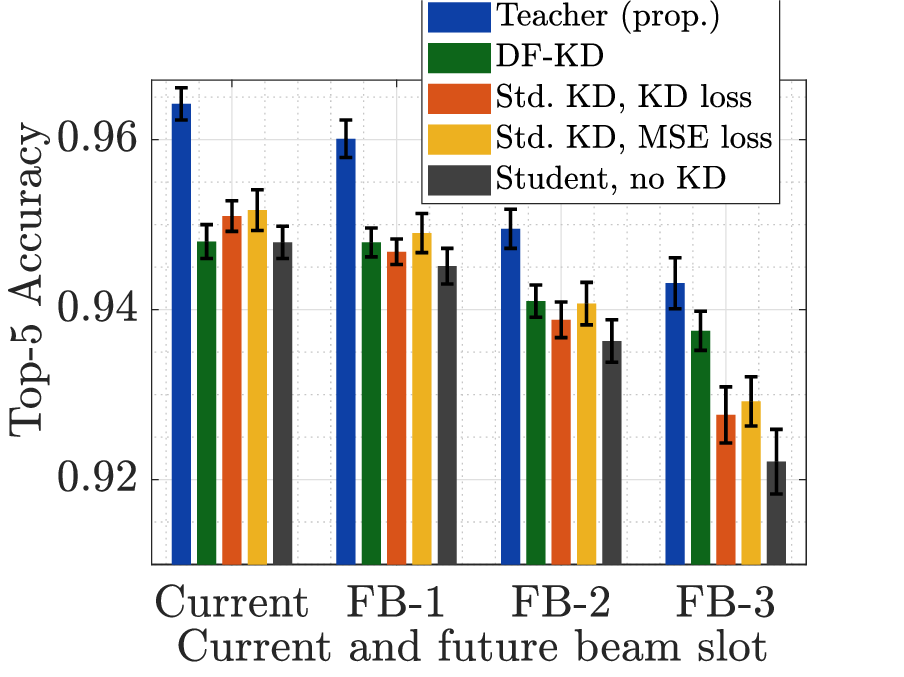 }
\label{fig_dfvskd_top5}
}
\caption{Top-1 and Top-5 accuracies of the teacher, the proposed DF-KD student, standard KD students, and the student 
trained without~KD, i.e., ``student no KD", where for the KD loss, $\alpha=0.6$ and $T=2$. } 
\label{fig_kdvsdfkd}
\vspace{-3 mm}
\end{figure}

\textit{Standard KD vs.\ DF-KD performance:} Fig.~\ref{fig_kdvsdfkd} 
presents the Top-1 and Top-5 accuracies of the proposed teacher 
and various KD configurations. We consider the student trained 
without KD as a lower bound, and two standard KD variants, using the KL loss and the MSE loss, as oracle upper bounds. This is because they are trained using real LiDAR data. 
The standard KD loss is given by:
\begin{equation}
    \mathcal{L}_{\text{KD}} = \alpha \mathcal{L}_{\text{KL}} + 
    (1 - \alpha)\mathcal{L}_{\text{cross-ent.}},
    \label{eq_std_kd}
\end{equation}
where $\mathcal{L}_{\text{KL}}$ is the KL loss in~\eqref{eq_student_kl_loss} 
and ${\mathcal{L}_{\text{cross-ent.}} = -\sum_{i=1}^{M} y_{\text{true},i} 
\log q_i}$ is the cross-entropy loss, where $y_{\text{true},i}$ is 
the true label for class $i$ and $q_i$ is the student's predicted 
probability for beam $i$.

The results show that the proposed DF-KD framework consistently 
outperforms the student trained without KD across all the future beam prediction 
slots in both Top-1 and Top-5 accuracies, demonstrating the 
effectiveness of knowledge transfer. As 
expected, the standard KD variants, which have access to real 
training data, achieve higher accuracy than DF-KD, with the 
standard KD (KL loss) reaching $57.98\%$ Top-1 current-slot 
accuracy compared to $54.12\%$ for DF-KD. However, this gap 
narrows progressively at future beam slots, confirming that 
DF-KD provides a practically viable alternative when the original LiDAR training dataset is unavailable. Among the standard KD 
variants, the MSE and KL losses yield comparable performance, 
consistent with the findings in Fig.~\ref{fig_Limpact_}.

\textbf{Complexity Analysis:} The additional complexity 
introduced by LiDAR arises from two aspects: sensing 
overhead and computational cost. In terms of sensing 
overhead, each LiDAR frame produces a preprocessed 
feature vector of dimension $D = 216$. With an 
observation window of $L = 8$ frames and $V = 3$ 
zero-padded future slots, the full input tensor per 
inference is of size $11 \times 216 = 2{,}376$ values, 
requiring the BS to continuously buffer and preprocess 
$L$ consecutive LiDAR scans before each prediction 
step. In terms of computational cost, the proposed CNN-GRU teacher has $189{,}952$ trainable parameters and requires approximately $3.98 \times 10^6$ 
floating-point operations (FLOPs) per inference, whereas the DF-KD student has only $8{,}096$ parameters and 
$0.10 \times 10^6$ FLOPs, corresponding to reductions 
of $\sim$$23$ times in parameter count and 
$\sim$$38$ times in FLOPs relative to the teacher.
For reference, the GRU-based model of~\cite{Ahmed_LiDar_COML} has $96{,}640$ parameters, making 
the proposed student approximately $12$ times more compact even relative to this lighter baseline. These results confirm that the proposed DF-KD framework effectively addresses the computational burden of LiDAR-based beam tracking by producing a highly compact student model well-suited for resource-constrained vehicular 
deployment, without requiring access to the original LiDAR training data after the teacher training phase.


\section{Conclusions}
This paper proposed a data-free knowledge distillation (DF-KD) framework for LiDAR-aided mmWave beam tracking, 
eliminating the need for real training data after the 
initial teacher training phase. A knowledge inversion approach was adopted, where a generator was trained via 
a metadata loss to synthesize LiDAR-like sequences that 
preserved the teacher's internal feature distribution. 
A CNN-GRU teacher architecture was introduced to provide 
a more expressive knowledge source, and both KL and MSE losses were considered for student~training.

Simulation results confirmed the effectiveness of the 
proposed framework. The metadata loss proved essential 
for informative synthetic data generation, the CNN-GRU teacher consistently outperformed GRU-only alternatives in the data-free setting, and the MSE loss offered a 
simpler alternative to standard KD with fewer 
hyperparameters. Future work will investigate extending the framework to NLoS environments, broader speed 
conditions, and multimodal sensing scenarios.

\bibliographystyle{ieeetr}
\bibliography{Bib_References/conf_short,
Bib_References/IEEEabrv,
Bib_References/Bibliography, Bib_References/ML_Bio, Bib_References/multimodalsensing_Bio }

\end{document}